\documentclass[%
 reprint,
superscriptaddress,
 amsmath,amssymb,
 aps,
prb,
]{revtex4-2}

\usepackage{graphicx}% Include figure files
\usepackage{dcolumn}% Align table columns on decimal point
\usepackage{bm}% bold math
\usepackage{booktabs}
\usepackage[hidelinks]{hyperref}
\usepackage{float}
\usepackage{multirow}
\usepackage{tcolorbox}

\usepackage[caption=false]{subfig}
\usepackage[mathlines]{lineno}
\usepackage{siunitx}

\usepackage{cleveref}

\begin{document}

%\preprint{APS/123-QED}

\title{A Two-Regime Statistical Framework for Wind-Power Distributions: From Wind-Speed Fluctuations to Turbine Control}% Force line breaks with \\
\author{S. Mitra}
\affiliation{Nonlinear and Non-equilibrium Physics Unit, OIST Graduate University, 1919-1 Tancha, Onna, Okinawa, 904-0495, Japan}
\author{S. E. Lakhal}
\affiliation{Nonlinear and Non-equilibrium Physics Unit, OIST Graduate University, 1919-1 Tancha, Onna, Okinawa, 904-0495, Japan}
\affiliation{Center for Mathematical Morphology (CMM), MINES ParisTech, 35 rue Saint Honoré, 77305, Fontainebleau, France}
\author{C. P. Connaughton}
\affiliation{London Mathematical Laboratory, 8 Margravine Gardens, London W6 8RH, UK}
\author{J. E. Sardonia}
\affiliation{Exus Renewables North America, Pittsburgh, Pennsylvania, USA}
\author{M. M. Bandi}
\email[]{bandi@oist.jp}
\affiliation{Nonlinear and Non-equilibrium Physics Unit, OIST Graduate University, 1919-1 Tancha, Onna, Okinawa, 904-0495, Japan}

\date{\today}

\begin{abstract}

Wind-power variability is a major challenge for the reliable integration of utility-scale wind energy into modern power systems. Although wind-speed statistics are often described by simple parametric distributions, translating these statistics into turbine-level power fluctuations is nontrivial because the wind-speed–power relationship is highly nonlinear and changes across different turbine operating regimes. Here, we develop a two-regime statistical framework for wind-power distributions. In the aerodynamic operating regime, between the cut-in and rated speed, the turbine power follows the approximate cubic wind-speed relation. Starting from a physically motivated Rician model for the wind-speed magnitude, we derive an analytical expression for the corresponding wind-power distribution using a nonlinear change of variables. In the control-dominated near-rated regime, where active blade-pitch and generator control regulate the turbine output, the aerodynamic transformation is no longer applicable. Instead, we characterize the power deficit relative to the rated power and show empirically that its continuous tail is well described by a bounded stretched-exponential distribution for both individual turbines and wind-farm ensembles. These results provide a physically interpretable statistical description of wind-power fluctuations across the full operational range of utility-scale wind turbines.

\end{abstract}

                              
\maketitle

\section{Introduction}
\label{intro}

As the inclusion of renewable energy sources in modern power grids continues to increase \cite{9183858,DALALA2022123361}, maintaining grid stability becomes increasingly challenging \cite{chum2011ipcc,apt2014variable,schmietendorf2017impact,smith2022effect} due to the inherent variability of renewable generation, particularly from wind and solar power \cite{mackay2016sustainable,apt2007spectrum,bel2016grid,bandi2017spectrum,bel2019geographic,bel2024spectral,bandi2016variability}. High renewable penetration reduces the effective load-carrying capacity (ELCC) \cite{6963441,SONG2024131558} and rotational inertia of the grid \cite{ajami2026impact,bevrani2009robust,10413919,PADHAN2013242}, making power systems more susceptible to frequency fluctuations \cite{lim2022,ARYANI2022106091}, transient instabilities, and dynamic load-balancing challenges \cite{article_ingola,apt2014variable}.

Fluctuations in renewable generation also impose additional stress on electrical infrastructure and transmission networks \cite{menck2014dead,schmietendorf2017impact,rohden2012self}, increasing operational costs associated with grid regulation and ancillary services, including reserve generation, energy storage, and dissipative balancing mechanisms \cite{apt2014variable}. Achieving a reliable, cost-effective, and flexible integration of renewable energy therefore requires a detailed understanding of generation-side variability \cite{veers2019grand,van2016long,peinke2024energie}, from which efficient forecasting, mitigation, and control strategies can be developed.

Among renewable sources, wind energy exhibits particularly strong fluctuations due to the intermittent and turbulent nature of atmospheric wind fields \cite{mackay2016sustainable,apt2007spectrum,bel2016grid,lakhal2026collective}. Consequently, the instantaneous turbine power output, $p(t)$, varies over a broad range of temporal and spatial scales \cite{bel2016grid,bandi2017spectrum,lakhal2026collective}. Understanding the statistical properties of these fluctuations is important not only for short-term forecasting and reliability assessment, but also for quantifying rare high-power events that can affect turbine operation, transmission infrastructure, and overall grid stability \cite{chum2011ipcc,apt2014variable,schmietendorf2017impact,smith2022effect}.

Accurate statistical models that connect wind-resource variability to generated power are therefore essential for the reliable integration of utility-scale wind energy. Wind-speed statistics are commonly described using standard parametric distributions \cite{akdaug2010use,akdaug2009new,lencastre2024modeling,shi2021wind,WADI2023237,CELIK2006105,CARTA2009933,CARTA2007518,CELIK2003693,CHANG2011272,CHAURASIYA20182299,CHELLALI2012379}, including the Weibull, Rayleigh, Rician, Gamma, log-normal, and Gumbel distributions, as well as mixture models. Translating these wind-speed statistics into power-output statistics, however, is nontrivial because the turbine power curve is strongly nonlinear \cite{bel2016grid,bandi2017spectrum,lakhal2026collective} and comprises distinct operating regimes associated with the cut-in, rated, and cut-out wind speeds \cite{article_ghaffari,LYDIA2014452} (Fig.~\ref{fig:speed_vs_power}). Between the cut-in and rated wind speeds, the generated power is well approximated by the cubic aerodynamic relation \cite{bel2016grid,bandi2017spectrum,lakhal2026collective,apt2014variable} $p\propto v^3$, whereas near the rated power the turbine output is governed primarily by blade-pitch and generator control \cite{Johnson,Narayana,DOMINGUEZGARCIA20124994,10.3389/fenrg.2023.1181996}, causing the power to saturate and depart from the cubic scaling (Fig.~\ref{fig:speed_vs_power}).

In this work, we develop a statistical framework that treats these two operating regimes separately. For the aerodynamic operating regime, we derive an analytical wind-power distribution from a physically motivated Rician model [Eq.~\ref{rician_pdf}] \cite{Yang,9397402,Famoye01111995} of the wind-speed magnitude combined with the approximate cubic wind-speed–power [Eq.~\ref{speed_to_power}] relation. The resulting distribution [Eq.~\ref{nor_power_pdf}] provides a direct statistical link between atmospheric wind-speed fluctuations and turbine-level power variability while retaining the physical interpretability of the underlying wind-speed model.

For the control-dominated near-rated regime, where the aerodynamic transformation is no longer applicable, we introduce the power deficit relative to the rated power [Eq.~\ref{eq:power_deficit}] and characterize its statistical behavior empirically [Eq.~\ref{eq:st_expo_ccdf}], by analyzing individual turbines as well as wind-farm ensembles. Together, these two complementary descriptions provide a unified statistical framework for wind-power fluctuations across the full operational range of utility-scale wind turbines.

We investigate the validation of the models using high-resolution operational data from four geographically distinct utility-scale wind farms. The dataset consists of 10-minute averaged wind-speed (m/s) and power (kW) measurements following International Energy Commission standard 61400-12~\cite{IEC61400-12-1}, spanning multiple years and containing observations from between 52 and 120 turbines per wind farm. The main characteristics of the wind farms are summarized in Table~\ref{Table:1}.

\begin{table}[ht]
    \small
    \centering
    \caption{Characteristics of the wind farms.}
    \label{Table:1}
    \footnotesize
    \begin{tabular}{@{}cccc@{}}
    \toprule
    \textbf{Farm} & \textbf{Turbines ($N_{\mathrm{tur}}$)} & \textbf{Area (km$^2$)} & \textbf{Data span (years)} \\
    \midrule
    1 & 52  & 89.72 & 2.7 \\
    2 & 80  & 68.80 & 4.9 \\
    3 & 120 & 93.08 & 3.6 \\
    4 & 64  & 80.94 & 3.0 \\
    \bottomrule
    \end{tabular}
\end{table}

The remainder of the article is organized as follows. In section~\ref{sec:speed_power_model} we derive the wind power distribution model from wind speed distribution model. Section~\ref{sec:model_fit} outlines the estimation procedure and comparison metrics. Section~\ref{sec:cdf_ccdf} analyzes the cumulative distribution function (CDF) and complementary cumulative distribution function (CCDF) based description of the model. Finally in section~\ref{sec:near_rated_dist}, we discuss the statistics of the near rated power.

\section{From wind speed to wind power distribution}
\label{sec:speed_power_model}

\begin{figure}[t!] 
    \centering
    \includegraphics[width=0.5\textwidth]{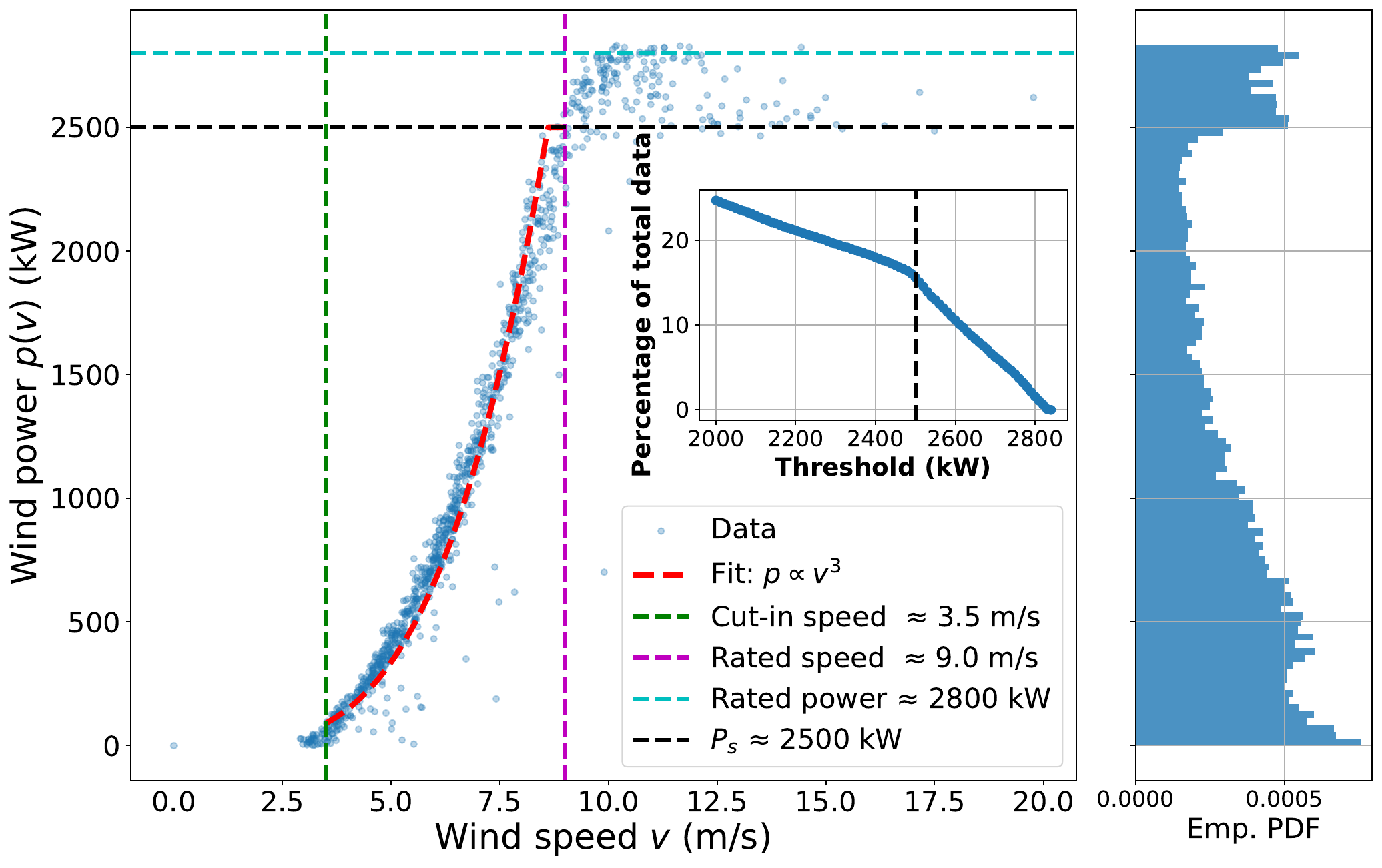}
    \caption{Empirical wind speed--power relation for a representative turbine in wind farm 1 using six months of data. (a) Main panel showing the cubic operating region and the near-rated saturation regime. (b) Right panel showing the empirical probability density function (PDF) of power. (c) Inset plot showing the percentage of total data within the near-saturated regime as a function of the threshold.}
    \label{fig:speed_vs_power}
\end{figure}

The power extracted by a wind turbine is determined by the kinetic energy flux of the incoming wind \cite{bel2016grid,bandi2017spectrum,lakhal2026collective,apt2014variable}. The relationship between the wind speed $v$ and the generated power $p(v)$ can be approximated by the following piecewise form \cite{article_ghaffari,LYDIA2014452} (Fig.~\ref{fig:speed_vs_power}):

\begin{equation}\label{eq:wind_speed_vs_power}
p(v)=
\begin{cases}
0, & v_{\min} \leq v < v_{\mathrm{cut\text{-}in}}, \\[6pt]
\dfrac{1}{2}\rho A C_p v^3 ,
& v_{\mathrm{cut\text{-}in}} \leq v \leq v_{\mathrm{rated}}, \\[8pt]
p_0-\epsilon, & v_{\mathrm{rated}} < v \leq v_{\mathrm{cut\text{-}out}},
\end{cases}
\end{equation}
where $\rho$ is the air density, $A$ is the rotor swept area, $C_p$ is the turbine power coefficient, $p_0$ is the rated power, and $\epsilon\ge0$ is a random deviation from the rated power arising from the turbine control and regulation mechanisms \cite{Johnson,Narayana,DOMINGUEZGARCIA20124994,10.3389/fenrg.2023.1181996}.

For wind speeds below the cut-in speed $v_{\mathrm{cut\text{-}in}}$, the available aerodynamic torque is insufficient to overcome the mechanical losses and inertia of the turbine, resulting in approximately zero power generation. In the operating region $v_{\mathrm{cut\text{-}in}}\le v\le v_{\mathrm{rated}}$, the generated power follows the well-known \cite{bel2016grid,bandi2017spectrum,lakhal2026collective,apt2014variable} cubic dependence $p(v)\propto v^3$. Once the wind speed exceeds the rated speed, blade-pitch and generator-torque control regulate the output to protect the turbine from excessive mechanical loading \cite{Johnson,Narayana,DOMINGUEZGARCIA20124994,10.3389/fenrg.2023.1181996}. Consequently, the power no longer follows the cubic law but instead enters a near-rated regime centered around the rated power $p_0$, producing a narrow distribution of power values below $p_0$ rather than a single constant value \cite{wang2021wind,subuh2025hybrid,veena2020artificially}.

Figure~\ref{fig:speed_vs_power} shows the empirical speed–power relation for a representative turbine in wind farm 1. The cut-in speed is approximately $v_{\mathrm{cut\text{-}in}}\approx3.8~\mathrm{m/s}$, the rated speed is $v_{\mathrm{rated}}\approx9~\mathrm{m/s}$, the cut-out speed is $v_{\mathrm{cut\text{-}out}}\approx15~\mathrm{m/s}$, and the rated power is $p_0\approx2800~\mathrm{kW}$. The near-rated regime begins at approximately $p_s \approx 2500~\mathrm{kW}$, corresponding to deviations $\epsilon$ of roughly $300~\mathrm{kW}$. Depending on the turbine control strategy, the wind conditions, and the observation period, approximately $15{\%}$–$30{\%}$ of the recorded power values lie within this regime (in the range of values approximately $100-300~\mathrm{kW}$), although the exact fraction varies among turbines and wind farms (see Table~\ref{Table:1}) (\Cref{fig:speed_vs_power,fig:speed_power_2,fig:speed_power_3,fig:speed_power_4}).
Note for the other wind farms considered in this study, the near-rated operating regime is comparatively narrow and primarily reflects the intrinsic turbine control that regulates power around the rated operating point. In contrast, Wind Farm 1 exhibits a substantially broader near-rated regime. According to the wind farm operator, this arises because the farm was subject to an additional facility-level power limit imposed by the grid interconnection agreement (for details see \hyperref[appendix1]{Appendix~A}). The wind speed--power curves for one representative turbine for the other three wind farms are also shown in \hyperref[appendix1]{Appendix~A}.

Since the cubic operating region is governed primarily by the stochastic properties of the wind speed, we first derive the power distribution for the interval $v_{\mathrm{cut\text{-}in}} \leq v \leq v_{\mathrm{rated}}$, where wind speed fluctuations are directly converted into power fluctuations through the approximate cubic relation $p(v)=Cv^3$. In contrast, the near-rated regime is dominated by turbine control and saturation mechanisms rather than the aerodynamic power law, and its statistical distribution is modeled separately.

Here, $C=\frac{1}{2}\rho A C_p$ collects the turbine-specific factors, including the air density, rotor swept area, and power coefficient [Eq.~\ref{eq:wind_speed_vs_power}]. Since $C$ only determines the characteristic power scale, it can be absorbed into the distribution parameters without affecting the functional form of the derived distribution. For notational simplicity, we therefore set $C=1$ and write

\begin{equation}
\label{speed_to_power}
p=v^3.
\end{equation}

To model the wind-speed statistics, we exploit the separation between the data sampling timescale,
$\tau_{\mathrm{sample}}=10~\mathrm{min}$,
and the turbine yaw-adjustment timescale,
$\tau_{\mathrm{yaw}}$,
which typically ranges from seconds to minutes \cite{article_starke,10.3389/fenrg.2021.626681,article_astolfi,article_zhang}
($\tau_{\mathrm{yaw}}\approx 1~\mathrm{min}$).
Since
$\tau_{\mathrm{sample}} \gg \tau_{\mathrm{yaw}}$,
the turbine continuously reorients itself toward the dominant incoming flow direction. Consequently, at each sampling instant, it is natural to decompose the instantaneous wind velocity into components parallel and perpendicular to the turbine-facing direction.

We therefore represent the instantaneous wind velocity at a fixed spatial point as a two-dimensional random vector,\[\vec{v}=(v_{\parallel},v_{\perp}),\]
where $v_{\parallel}$ and $v_{\perp}$ denote the velocity components parallel and perpendicular to the turbine-facing direction, respectively. We assume that these components are statistically independent and distributed as \cite{cui2016exact}
\[
v_{\parallel}\sim \mathcal{N}(\mu,\sigma^2),
\qquad
v_{\perp}\sim \mathcal{N}(0,\sigma^2).
\]
Here, $\mu$ represents the coherent mean-flow component aligned with the turbine orientation, while $\sigma$ characterizes the intensity of fluctuations around that mean flow. Both parameters are allowed to vary in time, for example across monthly windows, thereby capturing seasonal and environmental variability. 

The wind speed is defined as the magnitude of the velocity vector,

\begin{equation}
v=\sqrt{v_{\parallel}^2+v_{\perp}^2}.
\end{equation}
Under these assumptions, the wind-speed magnitude follows the Rician (Rice) distribution \cite{Yang,9397402,Famoye01111995},

\begin{equation}\label{rician_pdf}
f_R(v;\mu,\sigma)
=
\frac{v}{\sigma^2}
\exp\left(
-\frac{v^2+\mu^2}{2\sigma^2}
\right)
I_0\left(
\frac{v\mu}{\sigma^2}
\right),
\end{equation}
where $I_0$ denotes the modified Bessel function of the first kind of order zero \cite{book_bala}.

We now derive the corresponding wind-power distribution from the wind-speed distribution in the operational regime $
v_{\mathrm{cut\text{-}in}} \leq v \leq v_{\mathrm{rated}}$. Since this transformation between speed to power is monotonic over the interval of interest [Eq.~\ref{speed_to_power}], the power distribution follows directly from the standard change-of-variables formula \cite{book_bala},
\begin{equation}\label{ric_trans}
g(p;\mu,\sigma)
=
\int_{v_{\mathrm{cut\text{-}in}}}^{v_{\mathrm{rated}}}
dv\,
f_R(v;\mu,\sigma)\,
\delta(v^3-p),
\end{equation}
Substituting the Rician wind-speed distribution [Eq.\ref{rician_pdf}] yields
\begin{equation}\label{rician_power}
g(p;\mu,\sigma)
=
\frac{1}{3\sigma^2 p^{1/3}}
\exp\left[
-\frac{p^{2/3}+\mu^2}{2\sigma^2}
\right]
I_0\left(
\frac{p^{1/3}\mu}{\sigma^2}
\right),
\end{equation}

The derived distribution is valid over the bounded interval $p_c\leq p\leq p_s$, where $p_c$ corresponds to the cut-in power and $p_s$ denotes the onset of the near-rated saturation regime, beyond which the cubic aerodynamic power law no longer provides an adequate description of the turbine output. For the representative turbine shown in Fig.\ref{fig:speed_vs_power}, the onset of saturation occurs at approximately $p_s\approx2500\mathrm{kW}$. Within the interval $[p_c,p_s]$, the nonlinear transformation from wind speed to power substantially broadens the distribution and gives rise to a stretched-exponential-like decay [Eq.~\ref{ccdf_sexp}] \cite{article_Jakob,10.3389/fphy.2020.619729}. Although qualitatively similar power distributions can also be obtained by applying the same nonlinear transformation to other parametric wind-speed models \cite{akdaug2010use,akdaug2009new,lencastre2024modeling,shi2021wind,WADI2023237,CELIK2006105,CARTA2009933,CARTA2007518,CELIK2003693,CHANG2011272,CHAURASIYA20182299,CHELLALI2012379}, such as the Weibull or Gaussian distributions, the present formulation retains the physical interpretability of the underlying Rician wind-speed model. In particular, the parameters $\mu$ and $\sigma$ directly characterize the coherent mean-flow component and the fluctuation intensity of the wind field, allowing the power-distribution parameters to be interpreted in terms of physically meaningful wind dynamics rather than purely empirical fitting parameters.

Before fitting the model to empirical data, the PDF must be normalized over the physically accessible interval $[p_c,p_s]$. We therefore define the normalized conditional PDF as

\begin{equation}\label{nor_pdf}
g_{\mathrm{nor}}(p;\mu,\sigma)
=
\frac{g(p;\mu,\sigma)}
{\displaystyle \int_{p_c}^{p_s}g(p;\mu,\sigma)dp}.
\end{equation}
The normalization constant can be expressed in terms of the cumulative distribution function (CDF) associated with the power distribution $F_P(p)$, defined as,
\begin{equation}\label{CDF_def}
    F_P(p_a)= \mathbb{P}(p\leq p_a), 
\end{equation}
Since the mapping $p=v^3$ is monotonic and one-to-one over the domain of interest, the power CDF $F_P(p)$, can be written directly in terms of the wind-speed CDF $F_V(v)$ \cite{book_bala,rohatgi2015introduction},
\begin{equation}\label{speed_power_cdf}
F_P(p)
=
F_V\left(p^{1/3}\right),
\qquad
p_c\leq p\leq p_s.
\end{equation}
The normalized power PDF therefore becomes
\begin{equation}\label{nor_power_pdf}
g_{\mathrm{nor}}(p;\mu,\sigma)
=
\frac{g(p;\mu,\sigma)}
{
F_V(p_s^{1/3})
-
F_V(p_c^{1/3})
},
\qquad
p_c\leq p\leq p_s.
\end{equation}
For the Rician wind-speed distribution [Eq.~\ref{rician_pdf}], the corresponding wind-speed CDF is
\begin{equation}\label{speed_cdf}
F_V(v)
=
1-
Q_1\left(
\frac{\mu}{\sigma},
\frac{v}{\sigma}
\right),
\end{equation}
where $Q_1$ denotes the Marcum $Q$-function of order one \cite{article_colin,BARICZ2009265}.

\section{Model fitting and parameter estimation}
\label{sec:model_fit}

\begin{figure}[t!]
    \centering
    \includegraphics[width=0.5\textwidth]{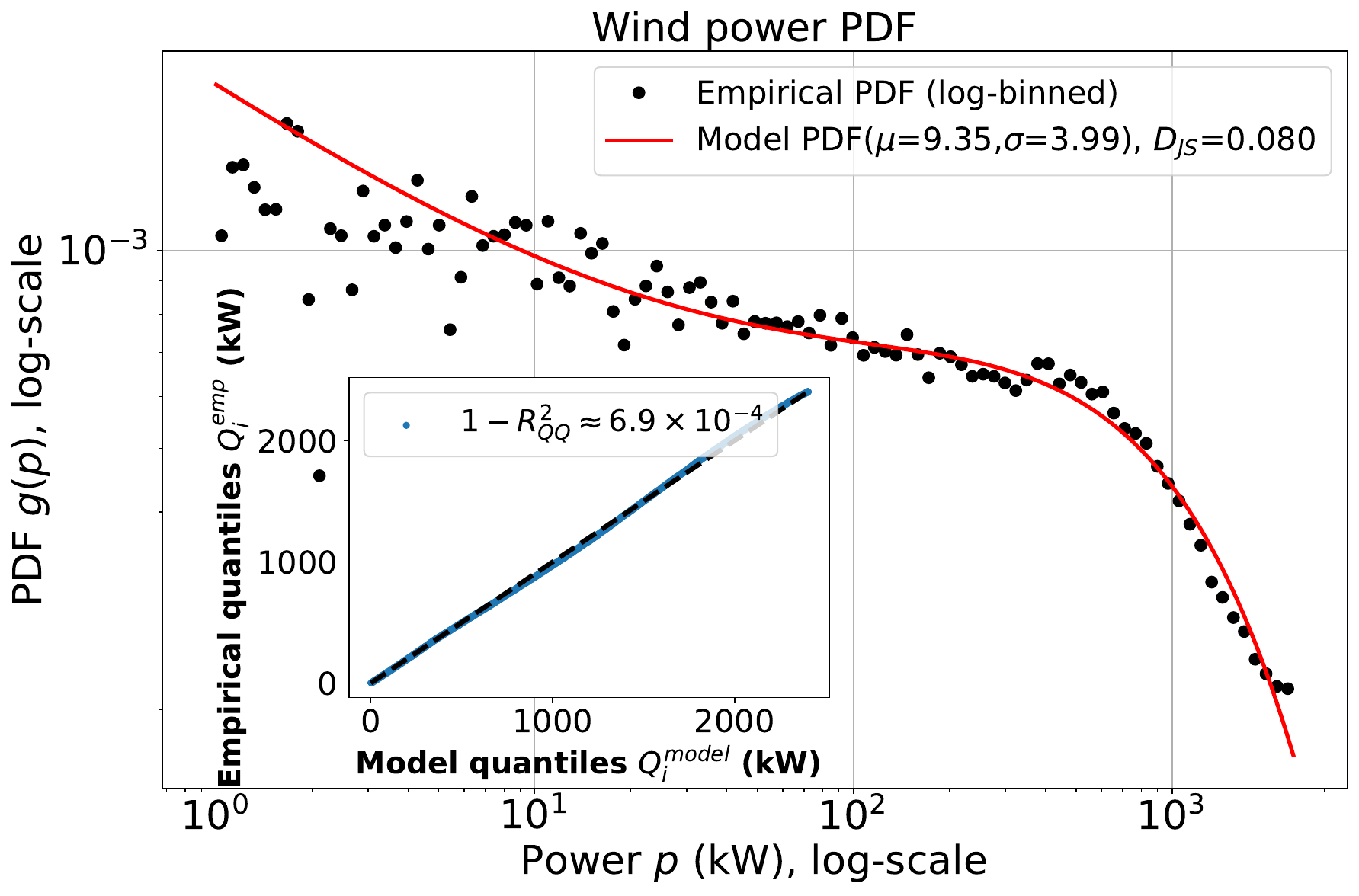}
    \caption{Empirical wind-power distribution for the representative turbine from wind farm 1 shown in Fig.~\ref{fig:speed_vs_power}. The distribution is plotted on a log--log scale using logarithmically binned histograms together with the fitted model distribution over the interval $p_c \le p \le p_s$. Inset shows the corresponding Q-Q plot.}
    \label{fig:power_pdf}
\end{figure}

The empirical wind-power distributions exhibit a broad upper tail. To estimate the model parameters, we employ maximum likelihood estimation (MLE), which infers the parameters directly from the raw power observations without requiring histogram binning. This avoids information loss and biases associated with the choice of bin width, particularly in the low- and high-power regions where observations are relatively sparse \cite{Newman_2005,Alstott_2014} (Fig.~\ref{fig:power_pdf}).

Within the MLE framework, the maximum-likelihood estimator is obtained by minimizing the negative log-likelihood (NLL) \cite{Farrell_Lewandowsky_2018},

\begin{equation}\label{mle_estimate}
\hat{\theta}= \arg\min_{\theta}
\left(
-\sum_{i=1}^{N}
\ln g_{\mathrm{nor}}(p_i|\theta)
\right),
\end{equation}
where $g_{\mathrm{nor}}(p_i|\theta)$ is the normalized model probability density function [Eq.~\ref{nor_power_pdf}] evaluated at the observation $p_i$, $\theta$ denotes the vector of model parameters, and $N$ is the total number of observations.

The estimated parameters for one representative turbine from each wind farm are summarized in Table~\ref{Table:MLE_params}, together with the corresponding $95\%$ confidence intervals \cite{Chang2025,monte_carlo} and goodness-of-fit measures.

\begin{table}[ht]
\centering
\caption{
Best-fit model parameters for representative turbines from the four wind farms.
The table also reports the corresponding $95\%$ confidence intervals and absolute goodness-of-fit measures. The consistently small values of the JS distance KS statistic, and logarithmic $L_2$ error indicate close agreement between the derived analytical model and the empirical power distributions. The  residual normalized variance of the QQ-plot, $1-R_{QQ}^2$,  is also reported here.
}
\label{Table:MLE_params}

\begin{tabular}{lcccc}
\toprule
Farm & 1 & 2 & 3 & 4 \\
\midrule

$\mu$ (m/s) 
& 9.32 & 9.61 & 10.36 & 9.21 \\

$95\%$ C.I. for $\mu$ 
& [9.28, 9.37] 
& [9.53, 9.69] 
& [10.29, 10.44] 
& [9.16, 9.27] \\

$\sigma$ (m/s) 
& 3.98 & 4.81 & 4.24 & 4.41 \\

$95\%$ C.I. for $\sigma$
& [3.94, 4.03] 
& [4.74, 4.89] 
& [4.19, 4.31] 
& [4.36, 4.48] \\

$D_{JS}$ 
& 0.081 & 0.084 & 0.087 & 0.113 \\

$D_{KS}$ 
& 0.0083 & 0.0128 & 0.0124 & 0.0146 \\

$L_2$ error 
& 0.026 & 0.033 & 0.027 & 0.0296 \\

$1-R^2_{QQ}$ 
& $6.5 \times 10^{-4}$ & $8.1 \times 10^{-4}$ & $7.5 \times 10^{-4}$ & $1.8 \times 10^{-3}$ \\

\bottomrule
\end{tabular}
\end{table}

Figure~\ref{fig:power_pdf} shows the empirical power distribution for the representative turbine from wind farm 1 presented in Fig.\ref{fig:speed_vs_power}, together with the fitted model distribution in log–log scale. The model is fitted over the interval $(p_c\leq p \leq p_s)$, corresponding to the operating regime in which the approximate cubic relation $(p\propto v^3)$ is valid [Eq.~\ref{speed_to_power}]. Near the rated power $p_0$, the empirical distribution develops a sharp accumulation due to turbine control saturation and power regulation (\Cref{fig:speed_vs_power,fig:speed_power_2,fig:speed_power_3,fig:speed_power_4}). Since this feature is not described by the continuous model, it is excluded from the fitting procedure.

To quantify the agreement between the empirical and model probability density functions (PDFs), we employ the Jensen--Shannon distance (JSD), a bounded and symmetric measure derived from the Jensen--Shannon divergence \cite{MENENDEZ1997307,hoyososorio2024representationjensenshannondivergence}. The distance is defined as
\begin{equation}\label{jsd}
D_{\mathrm{JS}}
\!\left(
g^{\mathrm{emp}},
g^{\mathrm{model}}
\right)
=
\sqrt{
JS\!\left(
g^{\mathrm{emp}}
\middle\|
g^{\mathrm{model}}
\right)
},
\end{equation}
where the Jensen--Shannon divergence is
\begin{equation}
JS\!\left(
g^{\mathrm{emp}}
\middle\|
g^{\mathrm{model}}
\right)
=
\frac{1}{2}
{\mathrm{KL}}
\!\left(
g^{\mathrm{emp}}
\middle\|
m
\right)
+
\frac{1}{2}
{\mathrm{KL}}
\!\left(
g^{\mathrm{model}}
\middle\|
m
\right),
\end{equation}
with
\[
m=\frac{g^{\mathrm{emp}}+g^{\mathrm{model}}}{2},
\]
and \({\mathrm{KL}}\) denoting the Kullback--Leibler (KL) divergence \cite{Brunton_Kutz_2022,KORESHI2022149},
\begin{equation}\label{kl_div}
{\mathrm{KL}}(f_1\|f_2)
=
\sum_k
f_{1k}
\log_2
\left(
\frac{f_{1k}}{f_{2k}}
\right).
\end{equation}
By construction, the Jensen--Shannon distance is symmetric and bounded between 0 and 1 (using logarithms of base 2), with smaller values indicating closer agreement between the empirical and model distributions. As with other histogram-based measures, its numerical value depends to some extent on the choice of histogram binning.

The small Jensen-Shannon distance values reported in Table~\ref{Table:MLE_params} demonstrate that the derived power distribution reproduces the empirical histogram with high accuracy across all four wind farms.

In addition to the PDF comparison, Fig.~\ref{fig:power_pdf} also includes a Quantile–Quantile (Q–Q) plot, which provides a more sensitive visualization of agreement throughout the distribution, particularly in the tail region. To quantify this agreement, we compute the coefficient of determination for the Q–Q plot,

\begin{equation}
R_{QQ}^{2}= 1-\frac{\sum_{i=1}^{n}\left(Q_i^{\mathrm{emp}}-Q_i^{\mathrm{model}}\right)^2}{\sum_{i=1}^{n}\left(Q_i^{\mathrm{emp}}-\bar{Q}^{\mathrm{emp}}
\right)^2
},
\end{equation}
where $Q_i^{\mathrm{emp}}$ and $Q_i^{\mathrm{model}}$ denote the empirical and model quantiles respectively, and $\bar{Q}^{\mathrm{emp}}=\frac{1}{n}\sum_{i=1}^{n}Q_i^{\mathrm{emp}}$, $n$ is the total number of quantiles used in the comparison. For all representative turbines across the four wind farms, the Q–Q plots remain close to the identity line over the fitted interval $p_c\leq p \leq p_s$, with $R_{QQ}^{2}$ values consistently exceeding $0.99$. This indicates that the model accurately reproduces both the bulk and tail behavior of the empirical power distribution within the domain where the cubic speed–power approximation is valid.

Along with the JS distance, Table~\ref{Table:MLE_params} also reports the Kolmogorov–Smirnov (KS) statistic [Eq.\ref{ks_dist}] and the $L_2$ error [Eq.\ref{l2_dist_tail}], whose definitions are based on the cumulative distribution function (CDF) [Eq.\ref{CDF_def}] and complementary cumulative distribution function (CCDF) [Eq.\ref{CCDF_def}] representations discussed in the following section. The consistently small values of these metrics further support the validity of the derived power-distribution model across all four wind farms.

\begin{figure}[t!]
    \centering
    \includegraphics[width=0.4\textwidth]{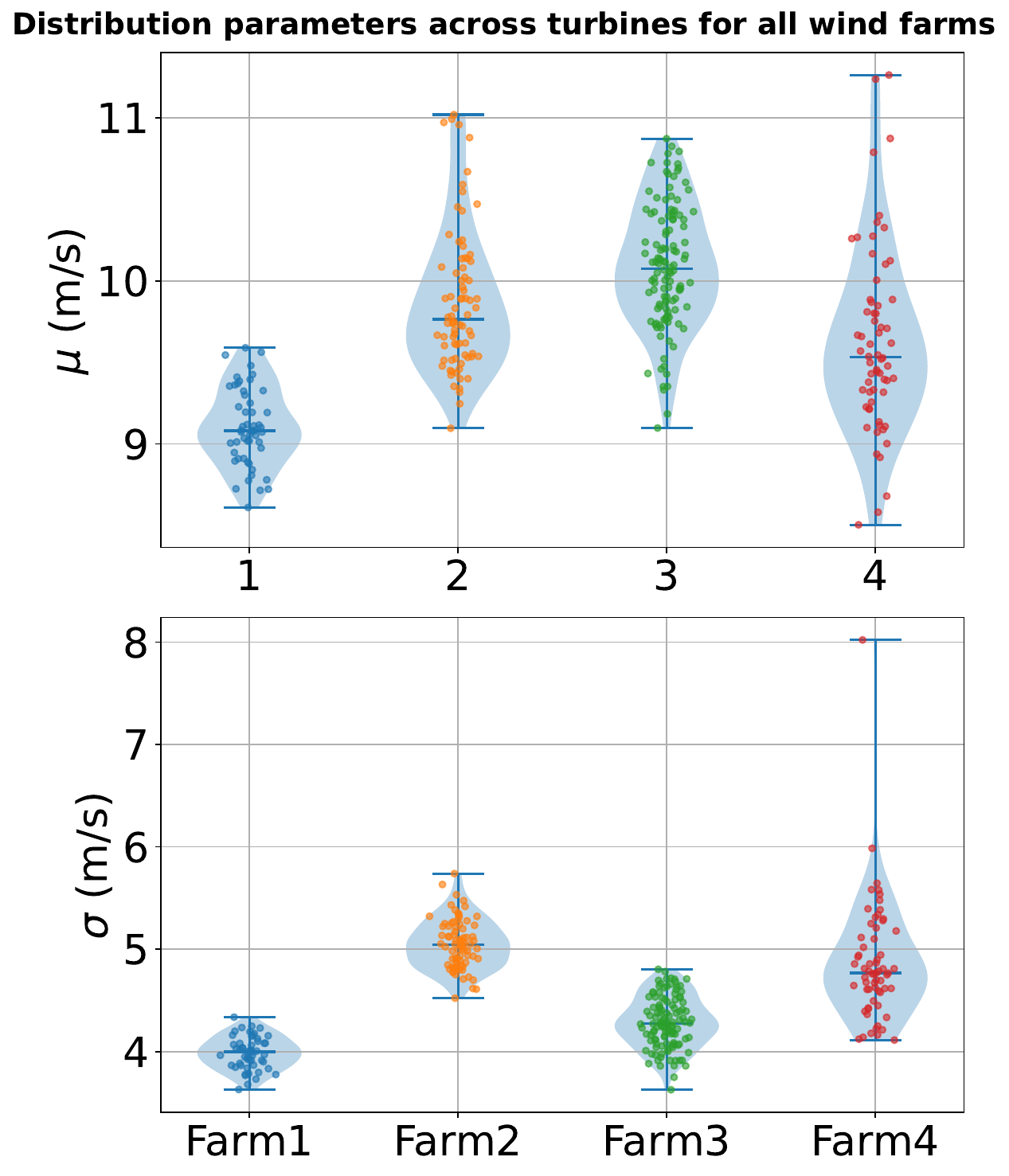}
    \caption{Distribution of mean-flow speed component $\mu$ (top panel) and deviation $\sigma$ parameters (bottom panel) for all turbines across the four wind farms (Table~\ref{Table:1}), shown as violin plots. The broad distributions indicate both intra-farm and inter-farm variability in the effective wind statistics and turbine operating conditions.}
    \label{fig:param_dist}
\end{figure}

Finally, Fig.~\ref{fig:param_dist} shows the distribution of the fitted model parameters for all turbines across the four wind farms using violin plots. The parameters exhibit both intra-farm and inter-farm variability, reflecting spatial heterogeneity in wind conditions as well as differences in turbine operating characteristics and control dynamics (see Table~\ref{Table:1}).

\section{CDF and CCDF analysis of wind-power distributions}
\label{sec:cdf_ccdf}

To characterize the behavior of the distribution across both low- and high-power regimes, we complement the probability density function (PDF) analysis with the cumulative distribution function (CDF) [Eq.~\ref{CDF_def}] and the complementary cumulative distribution function (CCDF) [Eq.~\ref{CCDF_def}], also known as the survival function \cite{book_majumdar,sabhapandit2019extremesrecords,book_bala,rohatgi2015introduction} . 

The CDF is particularly useful for analyzing the low-power regime and the bulk statistics of the distribution, while the CCDF provides a more stable characterization of the high-power tail \cite{book_majumdar,sabhapandit2019extremesrecords}, where direct PDF estimation becomes increasingly sensitive to finite sampling and histogram binning \cite{Newman_2005,Alstott_2014} (Fig~\ref{fig:power_pdf}).

Since the model is restricted to the physically relevant interval
$p_c\leq p\leq p_s$, the CDF defined through Eq.~\ref{CDF_def} must be normalized over this bounded domain. Using the one-to-one mapping between the wind-speed and wind-power within the interval $v_{\mathrm{cut\text{-}in}}\leq v\leq v_{\mathrm{rated}}$ [Eq.~\ref{speed_to_power}], the normalized power CDF $F^{nor}_P(p)$ can be written directly in terms of the wind-speed CDF $F_V(v)$ [Eq.~\ref{speed_power_cdf}] as
\begin{equation}\label{CDF_nor}
F^{nor}_P(p)
=
\frac{
F_V(p^{1/3})
-
F_V(p_c^{1/3})
}
{
F_V(p_s^{1/3})
-
F_V(p_c^{1/3})
},
\end{equation}
such that $F^{nor}_P(p_c) = 0$ and $F^{nor}_P(p_s) = 1$. The function $F_V(p^{1/3})$ is obtained from Eq.~\ref{speed_cdf}.

\begin{figure}[t!]
    \centering
    \includegraphics[width=1\linewidth]{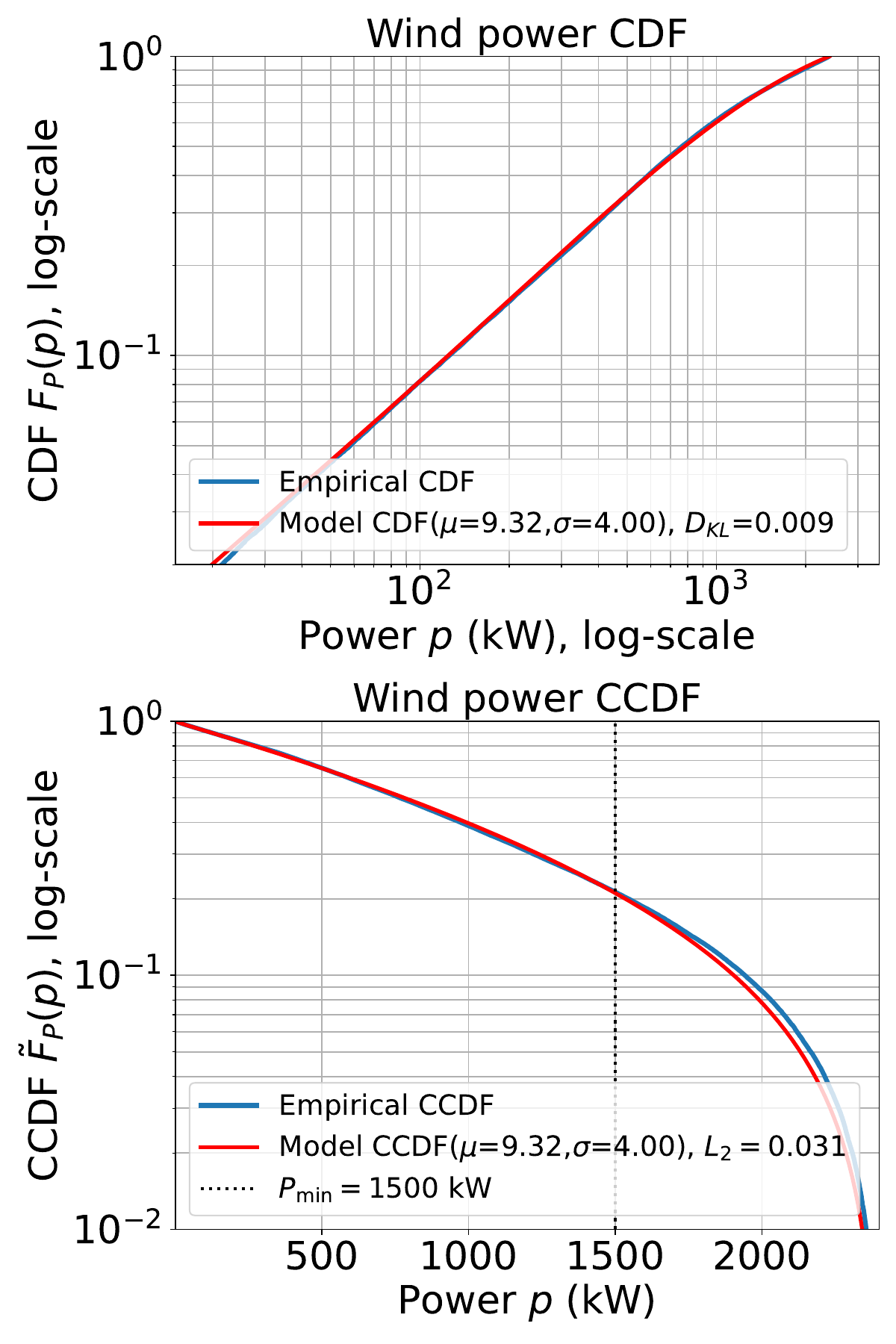}
    \caption{Empirical and fitted wind-power distributions for the representative turbine from wind farm 1 shown in Fig.~\ref{fig:speed_vs_power} in the operating regime: cumulative distribution function (CDF, top panel) and complementary cumulative distribution function (CCDF, bottom panel).}
    \label{fig:power_cdf_ccdf}
\end{figure}

Figure~\ref{fig:power_cdf_ccdf} shows (top panel) the empirical power CDF together with the fitted model CDF [Eq.~\ref{CDF_nor}] for the representative turbine from wind farm 1 shown previously in Figs.~\ref{fig:speed_vs_power} and \ref{fig:power_pdf}. The model CDF is evaluated using the maximum likelihood estimation (MLE) [Eq.~\ref{mle_estimate}] parameters listed in Table~\ref{Table:MLE_params}.

To quantify the agreement between the empirical and model cumulative distributions, we employ the Kolmogorov--Smirnov (KS) statistic \cite{Newman_2005,Alstott_2014},
\begin{equation}\label{ks_dist}
D_{\mathrm{KS}}
=
\sup_p
\left|
F^{\mathrm{emp}}(p)
-
F^{\mathrm{model}}(p)
\right|,
\end{equation}
where
$F^{\mathrm{emp}}(p)$
and
$F^{\mathrm{model}}(p)$
denote the empirical and model cumulative distribution functions (CDFs), respectively. The KS statistic measures the maximum absolute deviation between the two CDFs and is bounded between 0 and 1, with smaller values indicating closer agreement. Unlike the Jensen--Shannon distance [Eq.~\ref{jsd}], the KS statistic is computed directly from the empirical CDF and therefore does not require histogram binning, making it less sensitive to the choice of discretization. The corresponding KS values for representative turbines from each wind farm are reported in Table~\ref{Table:MLE_params}.

Finally, the high-power tail is analyzed using the complementary cumulative distribution (CCDF) function, or survival function, $\tilde{F}_P(p)$, defined as
\begin{equation}\label{CCDF_def}
\tilde{F}_P(p_a)=\mathbb{P}(p>p_a)=1-F_P(p_a).
\end{equation}
Using the asymptotic expansion of the power distribution [Eq.~\ref{rician_power}], it is easy to check that asymptotically CCDF function $\tilde{F}_P(p)$, can take stretched-exponential  \cite{article_Jakob,10.3389/fphy.2020.619729} shape:
\begin{equation}\label{ccdf_sexp}
    \tilde{F}_P(p) \sim \exp\left(-\frac{p^{2/3}}{2 \sigma^2}\right)
\end{equation}

Figure~\ref{fig:power_cdf_ccdf} displays (bottom panel) the empirical CCDF together with the fitted normalized model CCDF obtained from Eq.~(\ref{CDF_nor}) and Eq.~(\ref{CCDF_def}). 

To quantify the goodness of fit in the tail region, we employ a tail-restricted logarithmic $L_2$ distance between the empirical and model CCDFs. Unlike the JS distance [Eq.~\ref{jsd}] or KS statistic [Eq.~\ref{ks_dist}], which probe global agreement across the full distribution, the logarithmic $L_2$ measure emphasizes relative deviations in the rare-event tail. The metric is computed over the high-power region
$p_i\geq p_{\mathrm{tail}}$, with
$p_{\mathrm{tail}}=1500~\mathrm{kW}$:
\begin{equation}\label{l2_dist_tail}
L_2
=
\frac{1}{N_{\mathrm{tail}}}
\sum_{p_i\geq p_{\mathrm{tail}}}
\left[
\log \tilde{F}^{\mathrm{emp}}(p_i)
-
\log \tilde{F}^{\mathrm{model}}(p_i)
\right]^2.
\end{equation}
Here,
$N_{\mathrm{tail}}$
denotes the number of observations in the tail region,
$\tilde{F}^{\mathrm{emp}}(p_i)$
is the empirical CCDF, and
$\tilde{F}^{\mathrm{model}}(p_i)$
is the corresponding model CCDF. The resulting $L_2$ values for representative turbines from each wind farm are listed in Table~\ref{Table:MLE_params}.

Since the CCDF satisfies $0<\tilde{F}_P(p)<1$, the logarithmic discrepancy in Eq.~(\ref{l2_dist_tail}) remains well defined over the fitting interval, making $L_2$ a meaningful quantitative measure of the agreement between the empirical and model tails. In particular, $L_2=0$ corresponds to perfect agreement, while smaller values of $L_2$ indicate a closer match between the two CCDFs. The relatively small $L_2$ values obtained for all representative turbines therefore demonstrate that the proposed model accurately reproduces the high-power tail over the bounded interval $p_c \leq p \leq p_s$. Larger deviations are observed only in the vicinity of the rated power $p_0$, where turbine control saturation and mechanical power regulation \cite{article_ghaffari,LYDIA2014452} lead to an accumulation of observations near the cutoff, violating the assumptions underlying the continuous cubic-scaling approximation (\Cref{fig:speed_vs_power,fig:speed_power_2,fig:speed_power_3,fig:speed_power_4}).

\section{Distribution of the near-rated power region}
\label{sec:near_rated_dist}

The analytical power distribution derived in the previous section [Eq.~\ref{nor_power_pdf}] is valid only within the aerodynamic operating regime \cite{bel2016grid,bandi2017spectrum,lakhal2026collective,apt2014variable}, where the approximate cubic relation [Eq.~\ref{speed_to_power}] between wind speed and power holds. As the wind speed approaches the rated power $p_0$, active blade-pitch and generator-torque control regulate the turbine output to prevent excessive mechanical loading \cite{Johnson,Narayana,DOMINGUEZGARCIA20124994,10.3389/fenrg.2023.1181996}. Consequently, the generated power no longer follows the cubic aerodynamic law \cite{article_ghaffari,LYDIA2014452} [Eq.~\ref{speed_to_power}] but instead forms a pronounced accumulation \cite{wang2021wind,subuh2025hybrid,veena2020artificially} near the rated power (Fig.~\ref{fig:speed_vs_power}). Since this control-dominated regime cannot be described by the wind-speed transformation, it requires a separate statistical treatment.

To characterize fluctuations in the near-rated regime, we define the power deficit from Eq.~(\ref{eq:wind_speed_vs_power}) as,
\begin{equation}\label{eq:power_deficit}
\epsilon=p_0-p,
\end{equation}
where $p_0$ is the rated turbine power and $p$ is the measured power output. Thus, $\epsilon=0$ corresponds to operation at the rated power, while increasing $\epsilon$ represents progressively larger departures below the rated operating point.

For each turbine, the onset of the near-rated regime ($p_s$) was identified from the empirical speed--power curve (\Cref{fig:speed_vs_power,fig:speed_power_2,fig:speed_power_3,fig:speed_power_4}), typically lying $100$--$300~\mathrm{kW}$ below the rated power (see \hyperref[appendix1]{Appendix~A}). Only observations satisfying $p_s<p\le p_0$ were retained, corresponding to the bounded interval $0\le\epsilon\le\epsilon_{\max}$, where $\epsilon_{\max}=p_0-p_s$.

The empirical CCDF [Eq.\ref{CCDF_def}] of $\epsilon$ exhibits two distinct regimes. Close to $\epsilon=0$, a pronounced accumulation of observations arises from turbine control maintaining operation near the rated power. Beyond a transition value $\epsilon_t$, however, the CCDF displays a smooth, slowly decaying tail before terminating at the finite cutoff $\epsilon_{\max}$. Since the accumulation near $\epsilon=0$ is a consequence of turbine regulation rather than a continuously distributed stochastic process, continuous parametric distributions cannot simultaneously reproduce both features. We therefore model only the continuous component over the interval $\epsilon_t\le\epsilon\le\epsilon_{\max}$.

Motivated by the observed tail behavior, we model the complementary cumulative distribution (CCDF) [Eq.~\ref{CCDF_def}] \cite{book_majumdar,sabhapandit2019extremesrecords,book_bala,rohatgi2015introduction},
$\tilde{U}(\epsilon)=\mathbb{P}(\epsilon'>\epsilon)$,
using the bounded stretched-exponential \cite{article_Jakob,10.3389/fphy.2020.619729} form
\begin{equation}\label{eq:st_expo_ccdf}
\tilde{U}_{SE}(\epsilon,\lambda,\beta)
=
\frac{
e^{-(\lambda \epsilon)^{\beta}}
-
e^{-(\lambda \epsilon_{\max})^{\beta}}
}{
e^{-(\lambda \epsilon_t)^{\beta}}
-
e^{-(\lambda \epsilon_{\max})^{\beta}}
},
\end{equation}
where $\lambda$ is the rate parameter (with characteristic scale $1/\lambda$) and $\beta$ is the stretched-exponential shape parameter. The corresponding probability density, cumulative distribution, and maximum-likelihood estimation procedure are presented in \hyperref[appendix2]{Appendix~B}.

Figure~\ref{fig:rated_power_turbine} compares the empirical CCDF [Eq.\ref{CCDF_def}] of the power deficit for the representative turbine shown in Fig.~\ref{fig:speed_vs_power} with the fitted stretched-exponential model [Eq.~\ref{eq:st_expo_ccdf}]. The corresponding Kolmogorov--Smirnov distance [Eq.~\ref{ks_dist}], computed from the empirical CDF [Eq.\ref{CDF_def}] and the model CDF [Eq.~\ref{eq:st_exp_cdf}], demonstrates good agreement between the model and the observations.

To examine whether this behavior is universal across turbines within a wind farm, the near-rated observations from all turbines (see Table~\ref{Table:1}) were concatenated into a single dataset. Concatenation, rather than averaging, preserves the intrinsic variability because individual turbines experience different instantaneous wind conditions, whereas averaging would artificially suppress these fluctuations.

The resulting ensemble CCDF is shown in Fig.~\ref{fig:rated_power_ensamble}. The ensemble distribution closely matches the single-turbine results, indicating that the statistical properties of the power deficit are remarkably consistent across turbines within the same wind farm. Consequently, concatenating observations substantially improves the statistical sampling of the tail while preserving the bounded stretched-exponential form of the distribution. This suggests that the power-deficit distributions of individual turbines are approximately identically distributed, although not statistically independent owing to the strong correlation induced by the shared atmospheric forcing \cite{bandi2017spectrum,lakhal2026collective,apt2014variable}.

\begin{figure}[t]
    \centering
    \subfloat[]{%
        \includegraphics[width=1.\linewidth]{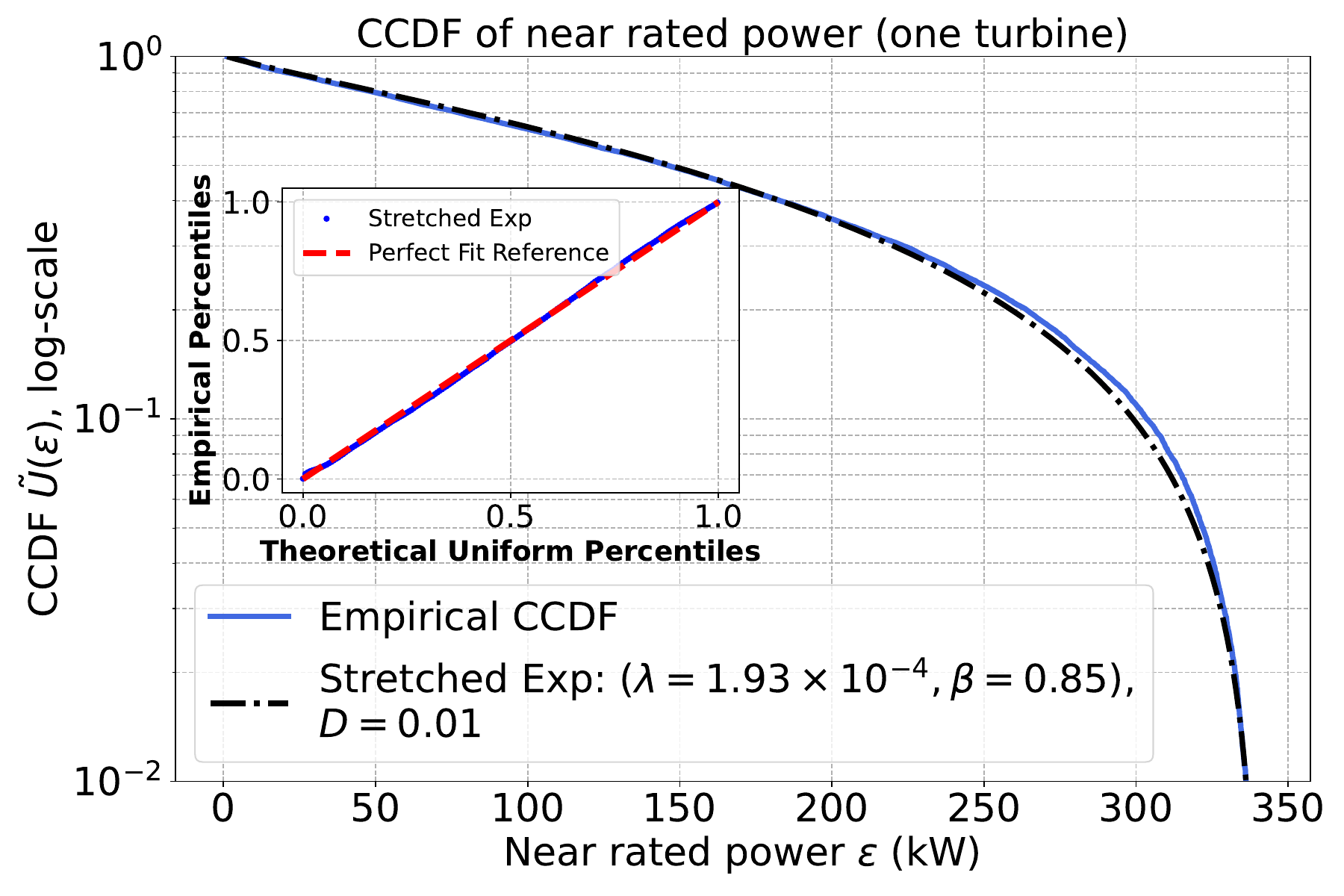}%
        \label{fig:rated_power_turbine}
    }
    \hfill
    \subfloat[]{%
        \includegraphics[width=1.\linewidth]{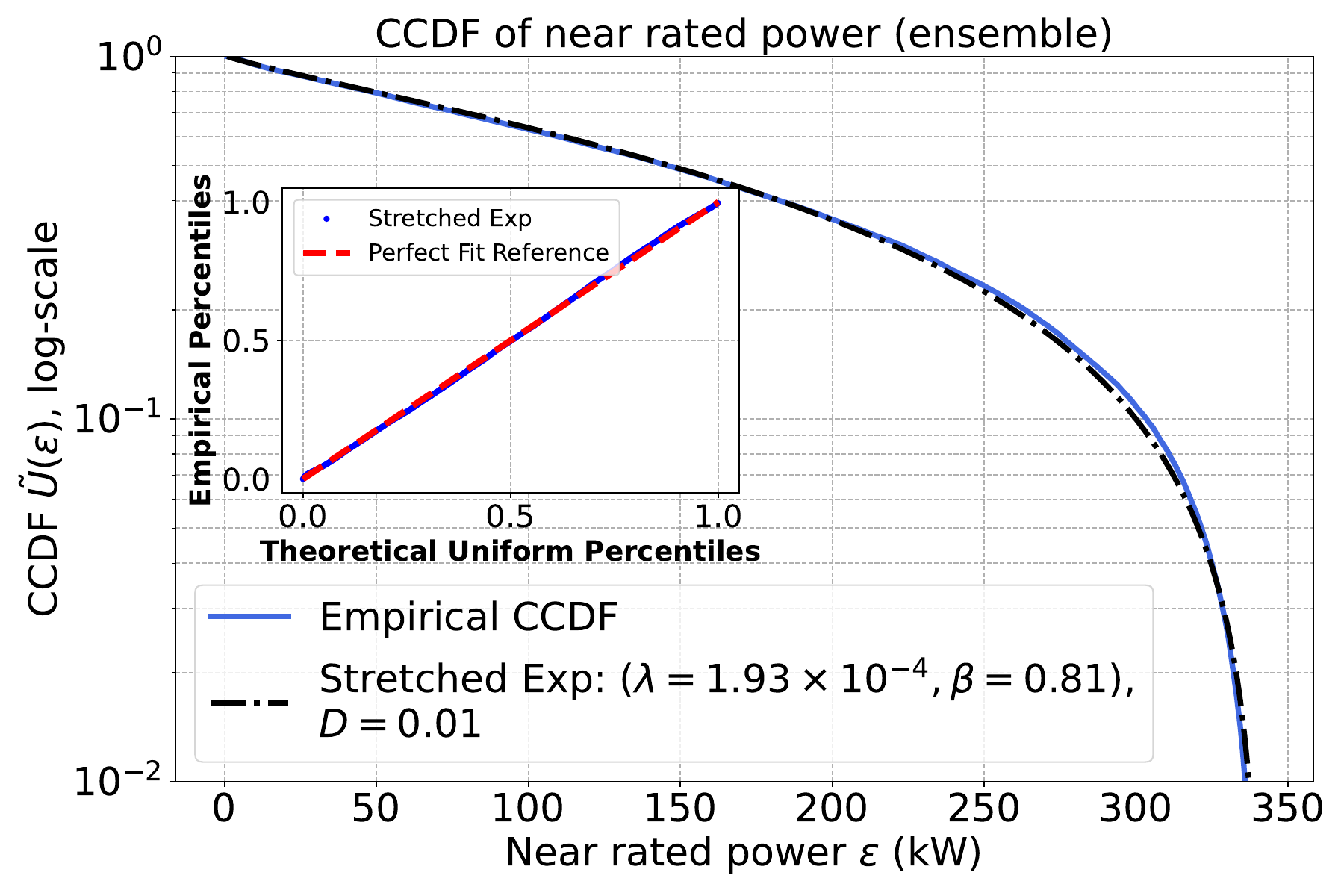}%
        \label{fig:rated_power_ensamble}
    }
    \caption{Empirical and model CCDF of the near-rated power deficit $\epsilon=p-p_{0}$ for (a) the wind farm 1 turbine shown in Fig.~\ref{fig:speed_vs_power} and (b) the wind farm 1 ensemble. Inset shows the corresponding probability--probability (P--P) plot, where theoretical percentiles are computed using the model CDF [Eq.~(\ref{eq:st_exp_cdf})] as $u_i=U_{\text{SE}}(\epsilon_{i},\lambda,\beta)$.}
    \label{fig:rated_power}
\end{figure}

\begin{table}[ht]
\centering
\caption{Maximum-likelihood estimates of the bounded stretched-exponential parameters for the near-rated power deficit. For each wind farm, the first group of columns corresponds to the representative turbine, while the second group corresponds to the ensemble obtained by concatenating the near-rated observations from all turbines within the farm. The Kolmogorov--Smirnov (KS) distance quantifies the agreement between the empirical and fitted distributions.}
\label{tab:near_rated_params}

\begin{tabular}{lccc ccc}
\toprule
&
\multicolumn{3}{c}{Representative turbine}
&
\multicolumn{3}{c}{Ensemble}
\\
\cmidrule(lr){2-4}
\cmidrule(lr){5-7}
Wind farm
&
$\lambda (\text{kW}^{-1})$
&
$\beta (-)$
&
$D_{\mathrm{KS}}$
&
$\lambda (\text{kW}^{-1})$
&
$\beta (-)$
&
$D_{\mathrm{KS}}$
\\
\midrule
Farm 1 & $1.93 \times 10^{-4}$ & 0.85 & 0.01 & $1.93 \times 10^{-4}$ & 0.81 & 0.01 \\
Farm 2 & $\le 10^{-5}$ & 0.34 & 0.01 & $ \le 10^{-5}$ & 0.76 & 0.04 \\
Farm 3 & $\le 10^{-5}$ & 0.27 & 0.02 & $\le 10^{-5}$ & 0.61 & 0.03 \\
Farm 4 & $7.44 \times 10^{-4}$ & 0.66 & 0.01 & $ 7.44 \times 10^{-4}$ & 0.67 & 0.01 \\
\bottomrule
\end{tabular}
\end{table}

Table~\ref{tab:near_rated_params} summarizes the maximum-likelihood estimates of the stretched-exponential parameters $(\lambda,\beta)$ for the representative turbine from each wind farm together with the corresponding ensemble estimates. For all wind farms, the fitted rate parameter $\lambda$ is very small, while for wind farms 2 and 3 the optimization reached the imposed lower bound ($\lambda\le10^{-5}$), indicating that the likelihood favors the small-$\lambda$ regime rather than a finite characteristic scale.

In the limit $\lambda\rightarrow0$, the bounded stretched-exponential CCDF [Eq.~\ref{eq:st_expo_ccdf}] reduces to the bounded power-law CCDF \cite{Newman_2005,Alstott_2014,clauset2009power,malevergne2005empirical},
\begin{equation}\label{eq:power_law_ccdf}
\tilde{U}(\epsilon,\beta)
=
\frac{\epsilon_{\max}^{\beta}-\epsilon^{\beta}}
{\epsilon_{\max}^{\beta}-\epsilon_t^{\beta}},
\end{equation}
whose corresponding probability density is the bounded Pareto distribution \cite{Newman_2005,Alstott_2014,clauset2009power,malevergne2005empirical},
\begin{equation}\label{eq:power_paw_pdf}
\xi(\epsilon,\beta)
\propto
\epsilon^{\beta-1},
\qquad
0<\beta<1.
\end{equation}

The consistently small estimates of $\lambda$ therefore suggest that the near-rated power-deficit statistics are primarily governed by the shape parameter $\beta$, with the characteristic scale $1/\lambda$ playing only a minor role over the observed range of the data. Consequently, for the wind farms considered here, the bounded stretched-exponential model is well approximated by its one-parameter bounded Pareto limit.

To verify this limiting behavior, we define the transformed empirical CCDF
\begin{equation}
\tilde{W}(\epsilon)
=
\frac{\epsilon_{\max}^{\beta}}
{\epsilon_{\max}^{\beta}-\epsilon_t^{\beta}}
-
\tilde{U}_{\mathrm{emp}}(\epsilon),
\label{eq:power_law_transform}
\end{equation}
where $\tilde{U}_{\mathrm{emp}}(\epsilon)$ is the empirical CCDF of the power deficit. In the bounded Pareto limit,
\begin{equation}
\tilde{W}(\epsilon)
=
\frac{\epsilon^{\beta}}
{\epsilon_{\max}^{\beta}-\epsilon_t^{\beta}}
\propto
\epsilon^{\beta},
\end{equation}
so that a log--log plot of $\tilde{W}(\epsilon)$ versus $\epsilon$ is expected to exhibit a straight line with slope $\beta$.

\begin{figure}[t!]
    \centering
    \includegraphics[width=1\linewidth]{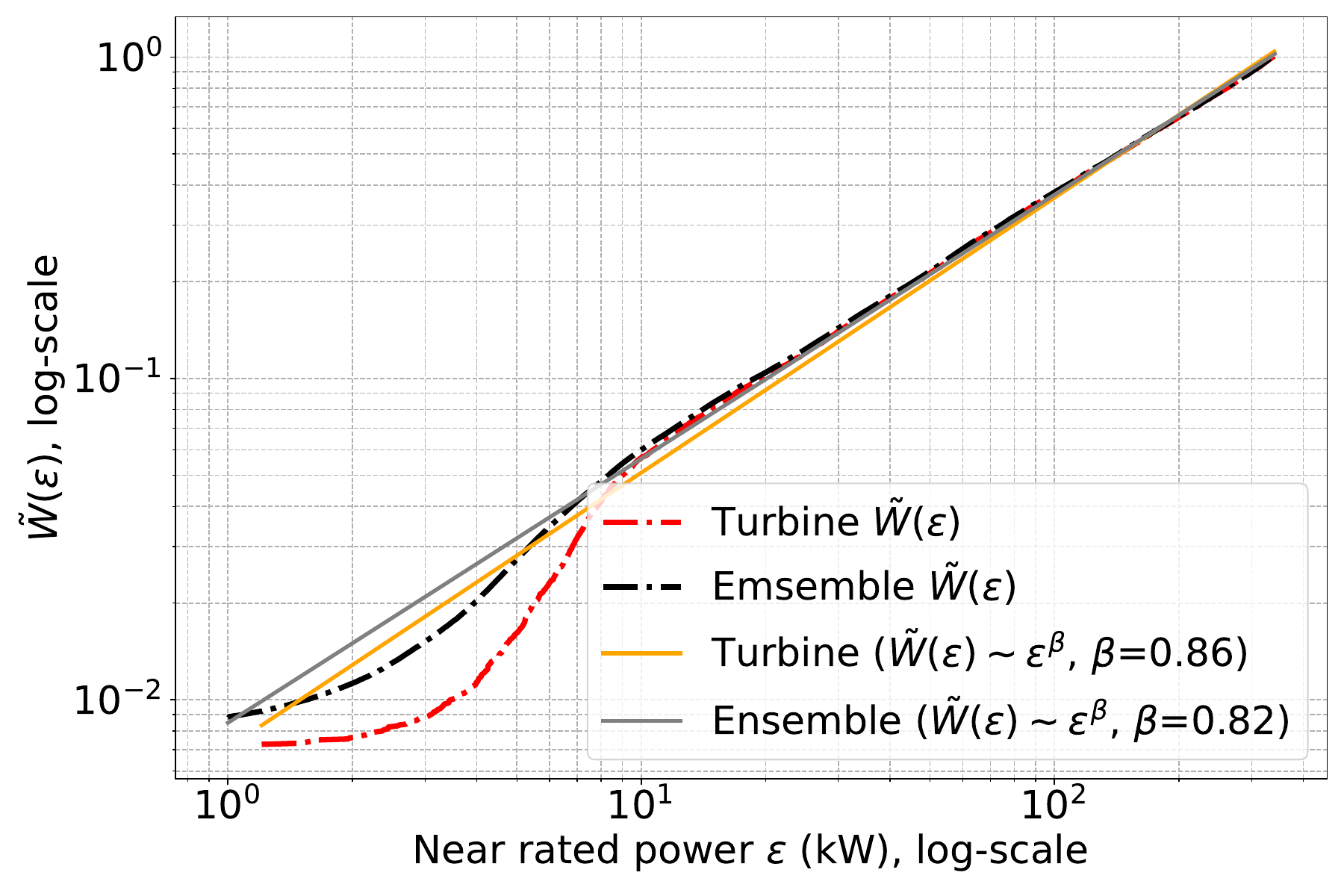}
    \caption{Transformed empirical CCDF $\tilde{W}(\epsilon)$ plotted against the power deficit $\epsilon$ on logarithmic axes for the representative turbine from wind farm 1 and the corresponding wind-farm ensemble. Solid lines denote power-law fits with exponent $\beta$ obtained from the slope of the log--log fit.}
    \label{fig:log_log_plot}
\end{figure}

Figure~\ref{fig:log_log_plot} shows the transformed empirical CCDF, $\tilde{W}(\epsilon)$, plotted against the power deficit $\epsilon$ on logarithmic axes for the representative turbine from wind farm 1 (Fig.~\ref{fig:rated_power_turbine}) and the corresponding wind-farm ensemble (Fig.~\ref{fig:rated_power_ensamble}). In both cases, the transformed CCDF exhibits an approximately two-decade power-law scaling. The measured slope is in excellent agreement with the exponent $\beta$ obtained independently from the maximum-likelihood fit of the bounded stretched-exponential distribution (Table~\ref{tab:near_rated_params}), thereby confirming the small-$\lambda$ bounded Pareto limit. Similar scaling behavior is observed for the remaining representative turbines and their corresponding wind-farm ensembles.

\section{Conclusion}
\label{sec:conclusions}

In this work, we developed and validated an analytical framework connecting wind-speed statistics to turbine-level wind-power statistics. Starting from a Rician model for the wind-speed magnitude [Eq.\ref{rician_pdf}], we derived an explicit expression for the wind-power distribution [Eq.\ref{nor_power_pdf}] within the aerodynamic operating regime between the cut-in speed and the onset of the near-rated saturation region. In this regime, the nonlinear cubic dependence of power on wind speed transforms the underlying wind-speed fluctuations into a stretched-exponential-like power distribution (Fig.\ref{fig:power_pdf}) [Eq.\ref{rician_power}].

The framework preserves the physical interpretability of the underlying Rician wind-speed model while explicitly incorporating the nonlinear transformation between wind speed and generated power. The parameters $\mu$ and $\sigma$ retain their direct physical meaning, representing the coherent mean-flow component and the intensity of wind-speed fluctuations, respectively. Through the cubic wind-speed–power relation, these physically interpretable parameters determine the statistical properties of the generated power, providing a direct connection between atmospheric variability and turbine-level power fluctuations.

Beyond the aerodynamic operating regime, the empirical power distributions exhibit a distinct control-dominated behavior near the rated power (\Cref{fig:speed_vs_power,fig:speed_power_2,fig:speed_power_3,fig:speed_power_4}) [Eq.~\ref{eq:wind_speed_vs_power}]. In this regime, active blade-pitch and generator control produce a pronounced accumulation of observations that cannot be explained by the cubic aerodynamic transformation alone. Introducing the power deficit, $\epsilon=p_0-p$, we showed that the continuous tail of the near-rated distribution is accurately described by a bounded stretched-exponential distribution [Eq.~\ref{eq:st_expo_ccdf}]. The fitted rate parameter is consistently very small, indicating that the data are well approximated by the small-$\lambda$ limit, in which the bounded stretched-exponential reduces to a one-parameter bounded Pareto distribution [Eq.~\ref{eq:power_paw_pdf}]. This behavior is consistently observed for both individual turbines (Fig.~\ref{fig:rated_power_turbine}) and wind-farm ensembles (Fig.~\ref{fig:rated_power_ensamble}).

The proposed two-regime statistical framework was validated using multiyear, 10-minute averaged measurements from four geographically distinct utility-scale wind farms containing between 52 and 120 turbines (Table~\ref{Table:1}). In the aerodynamic regime, the derived distribution accurately reproduces both the bulk and high-power tail of the empirical power distribution (Fig.~\ref{fig:power_cdf_ccdf}), with agreement quantified by the Jensen--Shannon distance [Eq.~\ref{jsd}], the Kolmogorov--Smirnov statistic [Eq.~\ref{ks_dist}], and the logarithmic CCDF-based $L_2$ metric [Eq.~\ref{l2_dist_tail}] (Table~\ref{Table:MLE_params}). In the near-rated regime, the bounded stretched-exponential (or equivalently its small-$\lambda$ bounded Pareto limit) accurately captures the continuous tail of the power-deficit distribution, as demonstrated by the empirical CCDFs and the corresponding Kolmogorov--Smirnov statistics (Table~\ref{tab:near_rated_params}).

This two-regime statistical framework provides a physically interpretable description of wind-power variability across the full operational range of utility-scale wind turbines and may serve as a useful basis for stochastic power forecasting, uncertainty quantification, and large-scale power-system modeling.

\appendix
\section{Wind speed--power curves for representative turbines from the remaining wind farms}
\label{appendix1}

\begin{figure}[t!]
    \centering
    \includegraphics[width=1.0\linewidth]{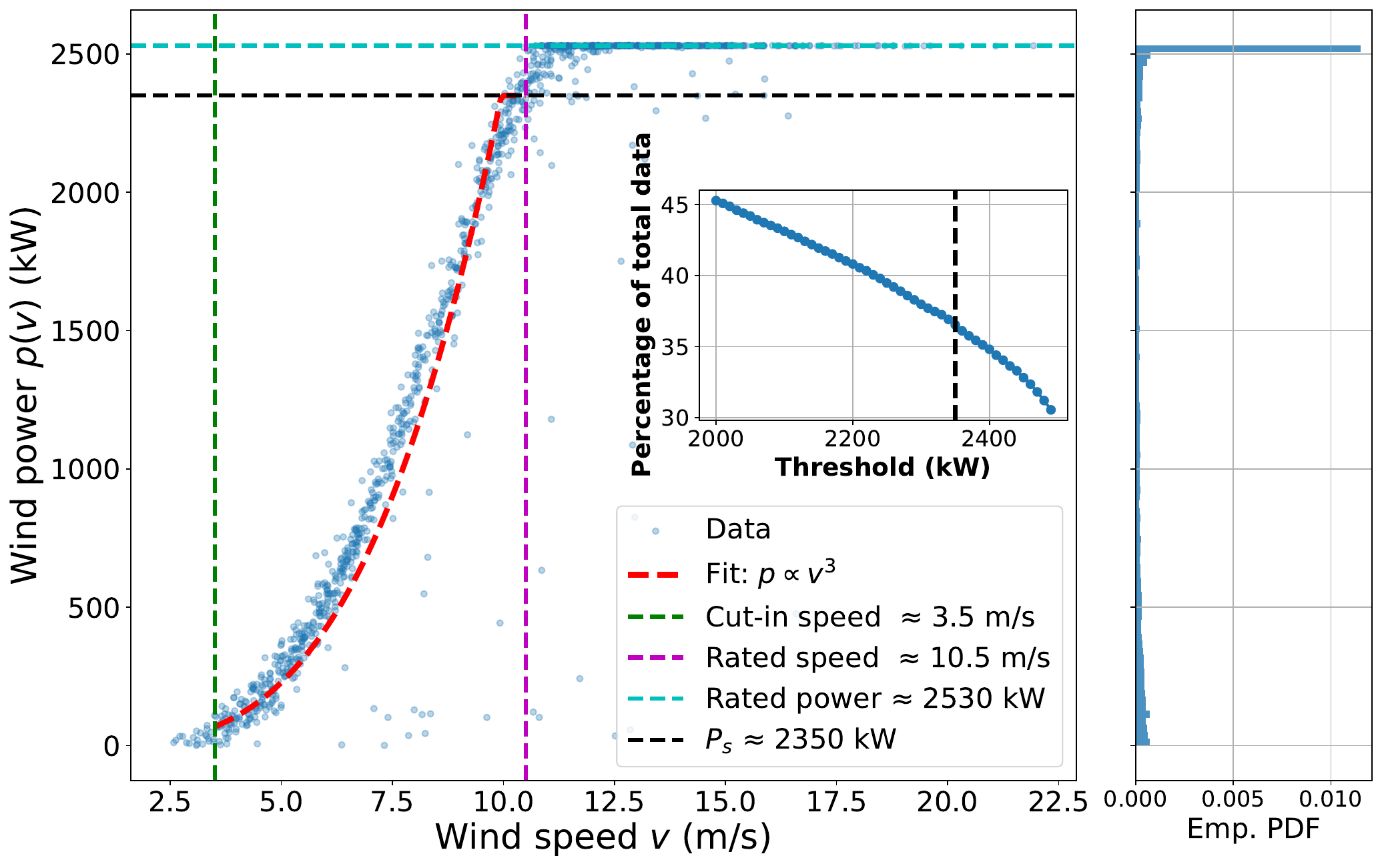}
    \caption{Empirical wind speed--power curve for a representative turbine from wind farm 2.}
    \label{fig:speed_power_2}
\end{figure}

\begin{figure}[t!]
    \centering
    \includegraphics[width=1.0\linewidth]{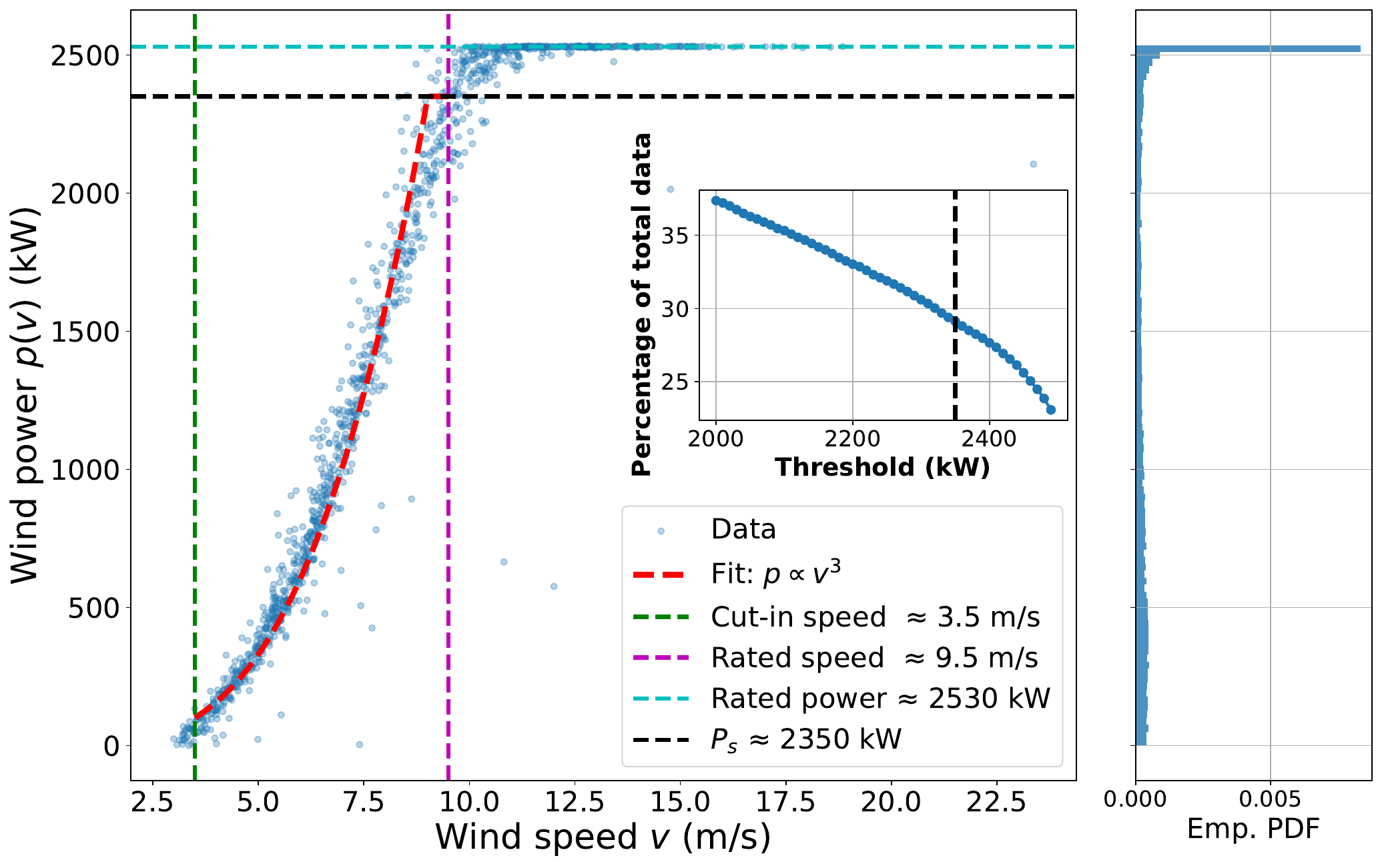}
    \caption{Empirical wind speed--power curve for a representative turbine from wind farm 3.}
    \label{fig:speed_power_3}
\end{figure}

\begin{figure}[t!]
    \centering
    \includegraphics[width=1.0\linewidth]{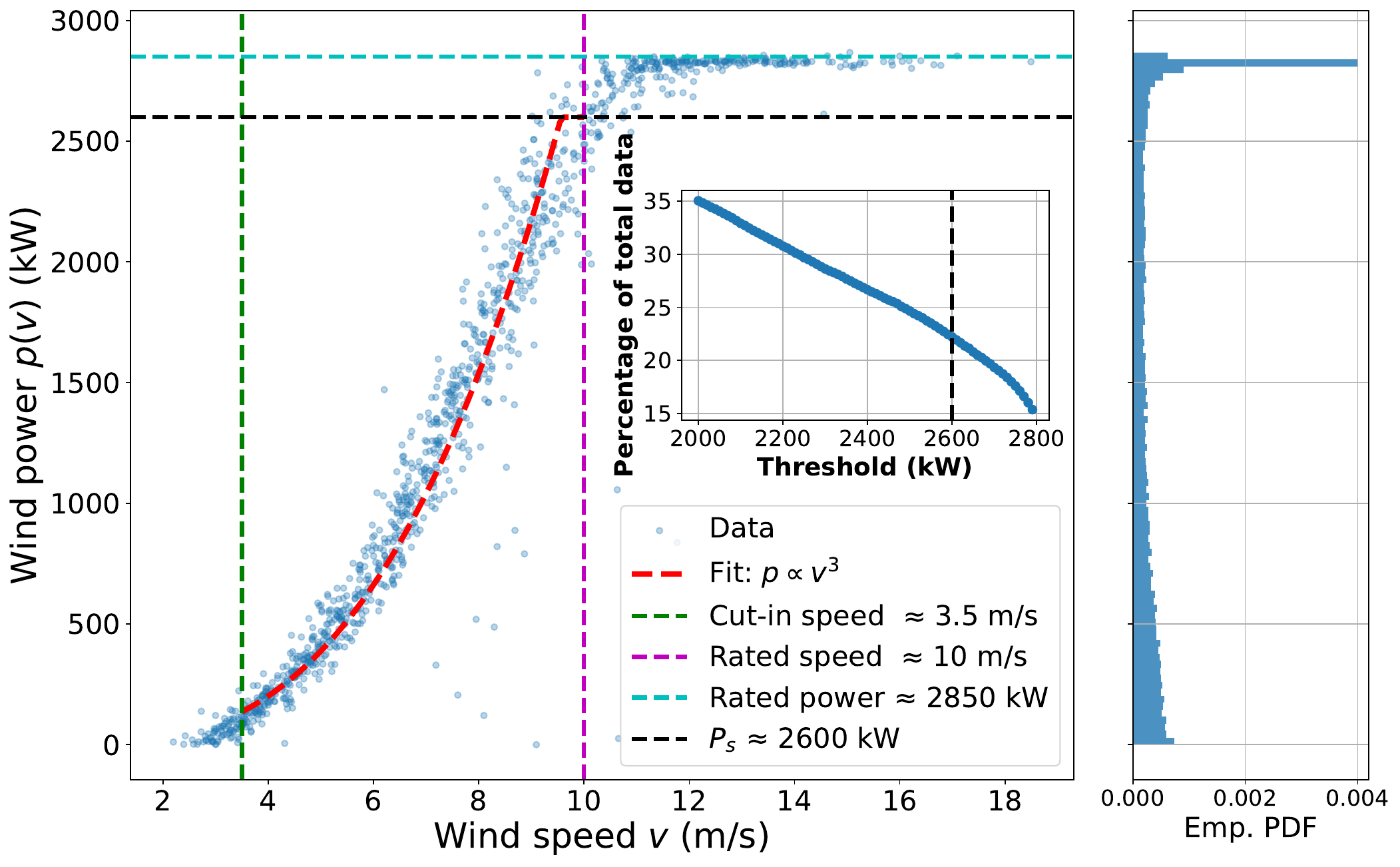}
    \caption{Empirical wind speed--power curve for a representative turbine from wind farm 4.}
    \label{fig:speed_power_4}
\end{figure}

Figure~\ref{fig:speed_power_2}--\ref{fig:speed_power_4} show the empirical wind speed--power curves for representative turbines from wind farms 2, 3, and 4. For each turbine, the cut-in speed $v_{\mathrm{cut\text{-}in}}$, rated speed $v_{\mathrm{rated}}$, rated power $p_0$, and the onset of the near-rated saturation regime $p_s$ are indicated. The aerodynamic operating regime and the control-dominated near-rated regime are also highlighted. The inset of each figure reports the fraction of observations (see Table~\ref{Table:1}) contained within the near-rated regime. Although the power curves exhibit small turbine-to-turbine variations within a given wind farm, they display similar overall characteristics, while systematic differences are observed between wind farms.

Importantly, for Wind Farm 1, the comparatively broad (Fig.~\ref{fig:speed_vs_power}) near-rated operating regime ($\epsilon_{\max}\approx300~\mathrm{kW}$) is explained by additional operational curtailment strategy rather than just by the intrinsic turbine characteristics. According to information provided by the wind farm operator, the total installed capacity of the farm (52 turbines, each rated at approximately  2800 kW) exceeded the maximum power permitted under the grid interconnection agreement. Consequently, once the total wind farm output approached the permitted facility limit, selected turbines were dynamically derated to prevent the aggregate generation from exceeding the allowable injection at the point of interconnection. Since the curtailment was not uniformly distributed among turbines, depending on factors such as maintenance outages and spatial variations in wind conditions, individual turbines operated over a range of approximately 2500–2800 kW even under rated wind speeds. This operational curtailment explains the broader near-rated power range observed for Wind Farm 1 compared with the other wind farms considered in this study.

\section{Bounded stretched-exponential distribution}
\label{appendix2}

To model the continuous tail of the near-rated power deficit, we employ bounded stretched-exponential distributions defined over the finite interval $\epsilon_t\le\epsilon\le\epsilon_{\max}$. This appendix summarizes the corresponding probability density functions, cumulative distribution functions, complementary cumulative distribution functions, and the maximum-likelihood estimation procedure used in the main text.

For an unbounded random variable $\epsilon\in[0,\infty)$, the stretched-exponential probability density function (PDF) is given by:
\begin{equation}
    \xi_{SE}(\epsilon;\lambda,\beta)= \beta \lambda (\lambda \epsilon)^{\beta-1}e^{-({\lambda \epsilon})^{\beta}}
\end{equation}
where $\lambda>0$ is the rate parameter and $0<\beta<1$ is the shape parameter..

The normalized conditional density in the bounded domain $\epsilon_t\le\epsilon\le\epsilon_{\max}$:
\begin{equation}
    \xi_{SE,nor}(\epsilon;\lambda,\beta)=\frac{\beta \lambda (\lambda \epsilon)^{\beta-1}e^{-(\lambda \epsilon)^{\beta}}}{e^{-(\lambda \epsilon _t)^{\beta}}-e^{-(\lambda \epsilon_{max})^{\beta}}}
\end{equation}

The corresponding CDF is:
\begin{equation}\label{eq:st_exp_cdf}
    U_{SE}(\epsilon;\lambda,\beta)=\frac{e^{-(\lambda \epsilon_t)^{\beta}}-e^{-(\lambda \epsilon)^{\beta} }}{e^{-(\lambda \epsilon_t)^{\beta}}-e^{-(\lambda \epsilon_{max})^{\beta}}}
\end{equation}

and the associated CCDF is:
\begin{equation}
    \tilde{U}_{SE}(\epsilon;\lambda,\beta)=1-U_{SE}(\epsilon;\lambda,\beta)=\frac{e^{-(\lambda \epsilon)^{\beta}}-e^{-(\lambda \epsilon_{max})^{\beta}}}{e^{-(\lambda \epsilon_{t})^{\beta}}-e^{-(\lambda \epsilon_{max})^{\beta}}}
\end{equation}

The parameters of model are estimated using the maximum-likelihood estimation (MLE) framework. The MLE is obtained by minimizing the negative log-likelihood (NLL),:

\begin{equation}\label{mle_estimate}
\hat{\theta}= \arg\min_{\theta}
l(\theta),
\end{equation}
Where $l(\theta)$ is the negative log-likelihood (NLL) function. where $\theta$ denotes the set of model parameters and the likelihood is evaluated over all observations within the fitting interval.

The resulting negative log-likelihood for the bounded stretched-exponential model is
\begin{equation}
\begin{aligned}
l_{SE}(\lambda,\beta)
&=
-\sum_{i=1}^{n}
\log
\xi_{SE,\mathrm{nor}}
(\epsilon_i;\lambda,\beta)
\\
&=
-n\log\beta
-
n\log\lambda
-
(\beta-1)
\sum_{i=1}^{n}
\log(\lambda\epsilon_i)
\\
&\qquad
+
\sum_{i=1}^{n}
(\lambda\epsilon_i)^\beta
+
n
\log
\left(
e^{-(\lambda\epsilon_t)^\beta}
-
e^{-(\lambda\epsilon_{\max})^\beta}
\right).
\end{aligned}
\end{equation}
Where $n$ is the total number of observations in the fitting interval.

\begin{acknowledgments}
SL was supported by a Japan Society for the Promotion of Science (JSPS) Postdoctoral Fellowship (grant no. P24714). MB was supported by JSPS KAKENHI (Grant No. 24KF0079). The data used in this work was obtained from Scout Clean Energy, Boulder CO.
\end{acknowledgments}

\bibliography{pop}

\end{document}